\documentclass{iau}
\usepackage{graphicx}
\usepackage{natbib}
\usepackage{hyperref}
\usepackage{caption}
\usepackage{subcaption}

\title[] 
{Wolf-Rayet stars: recent advances and persisting problems}

\author[Tomer Shenar]   
{Tomer Shenar$^1$       
 }

\affiliation{$^1$Anton Pannekoek Institute for Astronomy, Science Park 904, 1098 XH, Amsterdam, The Netherlands\\ email: {\tt T.Shenar@uva.nl}
}

\pubyear{2022}
\volume{361}  
\jname{Massive Stars Near and Far}
\editors{N.~St-Louis, J.~S.~Vink \& J.~Mackey, eds.}
\begin{document}

\maketitle

\begin{abstract}
Wolf-Rayet (WR) stars comprise a class of stars whose spectra are dominated by strong, broad emission lines that are associated with copious mass loss. In the massive-star regime, roughly 90\% of the known WR stars are thought to have evolved off the main sequence. Dubbed classical WR (cWR) stars, these hydrogen-depleted objects   represent a crucial evolutionary phase preceding core collapse into black holes, and offer a unique window into hot-star wind physics. Their formation is thought to be rooted in either intrinsic mass-loss or binary interactions. Results obtained from analyses using contemporary model atmospheres still fail to reconcile the derived properties of WR stars with predictions from stellar evolution. Importantly, stellar evolution models cannot reproduce the  the bulk of cWR stars, a problem that becomes especially severe at subsolar metallicity.  Next-generation model atmospheres  and upcoming observational campaigns to hunt for undetected companions promise a venue for progress.

\keywords{Wolf-Rayet stars, binaries:general}
\end{abstract}

\section{Introduction}
Wolf-Rayet (WR) stars are first and foremost a spectral class. Per definition, WR stars exhibit broad emission lines in lines that typically belong to helium, nitrogen for N-rich WR stars (WN), and carbon and oxygen for C- and O-rich WR stars (WC/WO).  The current census
of WR stars in our Milky Way Galaxy lists 667 objects \citep{vanderHucht2001, Crowther2007, Rosslowe2015}, but many thousands are known in resolved \citep[e.g.,][]{Breysacher1999, Neugent2018, Neugent2019} and unresolved \citep[e.g.][]{Hadfield2006, Kehrig2013, GomezGonzalez2021} extragalactic populations such as those of the Small and Large
Magellanic Clouds (SMC, LMC, Fig.\,\ref{fig:LMC_HRD}). 

The winds that characterise massive WR stars arise due to their proximity to the Eddington limit (i.e., high luminosity-to-mass, or $L/M$ ratio). On the one hand, proximity to the Eddington limit is reached by  very massive stars ($M_{\rm ini} \gtrsim 80-100\,M_\odot$) already on the main sequence due to their extreme luminosities \citep[e.g.,][]{deKoter1997, Rauw2004, Shenar2021}. On the other hand, massive stars that have evolved off the main sequence and have shed much of their hydrogen-rich layers \citep{Conti1976} launch powerful winds due to the mass reduction.
About 90\% of the known WR stars belong to the latter class of classical WR stars (cWR): hot ($T_{\rm eff} \gtrsim 30\,$kK), evolved (typically core He-burning), hydrogen-depleted massive stars. As the putative immediate progenitors of black holes, the properties and formation of cWR stars has enormous implications on the production of stellar-mass black holes, which are the seeds that produce the gravitational waves we are now regularly observing. 
These proceedings focus on cWR stars.

\section{Physical properties of cWR stars}

The analysis of WR spectra relies on sophisticated models that solve the radiative transfer problem in expanding atmospheres while relaxing the assumption of local thermodynamic equillibrium (non-LTE; \citealt{Hillier1991, Hamann2004}). Along with the mass-loss rate $\dot{M}$ and luminosity $L$, a central goal of such a fitting process is to derive the base effective temperature of the star $T_*$, which typically refers to the radial layer $r=R_*$ at which the mean Rosseland optical depth $\tau_{\rm Ross}$ reaches 10 or 20 (the difference is usually negligible). Hundreds of WR stars in our Galaxy and neighbouring galaxies have been thus analysed \citep{Hainich2014, Hainich2015, Shenar2016, Shenar2019, Hamann2019, Sander2020}. However, the resulting base temperatures are on average a factor of 2-3 lower than what is predicted from evolution models, a problem that became known as the WR "temperature problem" or, equivalently, the "radius problem".

What is "observed" in the spectrum are photons emitted from regions well above the stellar surface, which are closer to the photosphere ($\tau_{\rm Ross} \approx 2/3$, at $r = R_{2/3}$). While the derivation of $T_{\rm eff}(\tau = 2/3) = T_{2/3}$ is therefore relatively robust, the derivation of the base temperature $T_*$ fully depends on the adopted velocity law (see Lefever et al.\ in these proceedings). In the vast majority of analyses thus published, the velocity fields of the winds are fixed to the so-called $\beta$-law \citep{Castor1975}, typically adopting $\beta = 1$. 

Early investigations of the motion of clumps in the spectra of WR stars by \citet{Lepine1999} already suggested that $\beta=1$ laws do not yield a consistent solution to the outer velocity field (assuming the clumps trace the velocity field).  Moreover, 1D hydrodynamically-consistent models computed by \citet{Graefener2005} and, recently, by \citet*{Sander2020} show that the $\beta$-law is often a bad approximation. This is also seen in semi-analytical models provided by  \citet{Ro2016} and \citet{Poniatowski2021}, often revealing a "double-peaked" profile for the velocity field, where the first peak is associated with a dynamical inflation \citep[see also][]{Petrovic2006, Grassitelli2016, Graefener2012}. This dynamical inflation  may be the origin for the "temperature problem" of WR stars. However, it remains to be seen that such models can replicate the observed spectra of WR stars. 

And yet, 3D effects may also play an important role. First attempts at 3D modelling of the winds of WR stars have been recently published by \citet{Moens2022}. Among the plethora of rich results, an interesting finding is that the average velocity field in the 3D model no longer shows the "double-peaked" profile obtained in 1D models. However, this needs to be confirmed by future studies across the parameter regime. Surely, much progress and many insights will flow from this direction over the coming years.

Luckily, the measured luminosities of WR stars are much more robust. This is owing to the Stephan-Boltzmann equation, according to which $L \propto R_*^2 T_*^4 \approx R_{2/3}^2 T_{2/3}^4$ (neglecting the energy loss due to wind-driving). As the photospheric properties are generally well constrained, the luminosities are as well. As shown below, it is the luminosities of WR stars that are difficult to reconcile with evolution models.

\section{Formation and evolution}

Why should stars get rid of their hydrogen-rich envelopes? It was \citet{Paczynski1967} who proposed  that stripping by a binary companion could explain the formation of WR stars \citep[see also][]{Vanbeveren1998}. However, soon thereafter it was realised that massive stars can lose copious mass throughout various phases of their evolution, potentially involving eruptions during a Luminous Blue Variable (LBV) phase. Hence, stars exceeding a certain threshold mass may naturally evolve into the cWR phase, without any help from a companion. This became known as the Conti scenario, after Peter Conti who originally proposed this  \citep{Conti1976}.

\begin{figure}
\begin{center}
 \includegraphics[width=\textwidth]{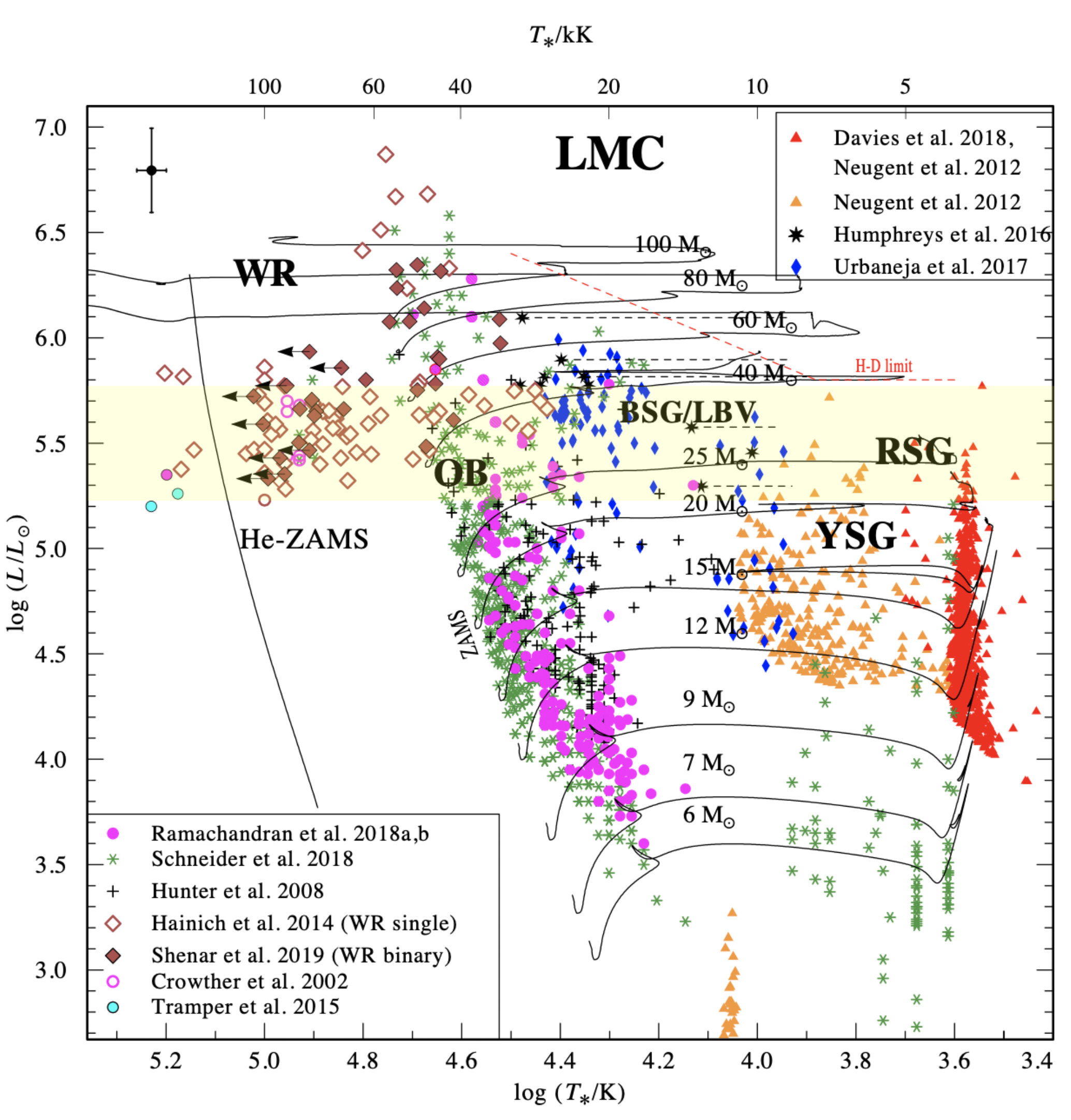} 
 \caption{Various stellar populations (see legend) shown on a Hertzsprung-Russell diagramm (HRD) of the LMC, adapted from \citet{Ramachandran2019} and:  \citet{Davies2018}, \citet{Neugent2012YSG}, \citet{Humphreys2016}, \citet{Urbaneja2017}, \citet{Schneider2018}, \citet{Shenar2019}, \citet{Hainich2014}, \citet{Crowther2002}, \citet{Tramper2015}, and \citet{Hunter2008}.  Tracks are single-star evolution tracks computed with the Modules for Experiments in Stellar Astrophysics (MESA) code \citep{Paxton2011},  taken from \citet{Gilkis2021}.  Tracks are single-star evolution tracks computed with the MESA code, adopted from \citet{Gilkis2021}.  The WN population (which comprises roughly 80\% of the total LMC WR population) is shown on the left part, with filled symbols depicting known binaries, and empty symbols depicting apparently-single WN stars.
The WR population is not reproduced, notably the apparently-single stars.  A similar picture is observed with other widely-used tracks, such as Geneva \citep{Eggenberger2021} and BPASS \citep{Eldridge2009, Eldridge2017}. Highlighted is the overlap between cool supergiants and the WR populations, which includes only the most luminous cool supergiants. Figure courtesy of Varsha Ramachandran and Avishai Gilkis.  }
  \label{fig:LMC_HRD}
\end{center}
\end{figure}

The question has not been settled ever since. In the late $20^{\rm th}$ century, intrinsic mass-loss generally took precedence, since mass-loss prescriptions published at the time, in combination with rotation, could broadly explain the census and observed ratios of the WR populations (e.g., WC/WN, WN/O, \citealt{Meynet2003}). However, in the past two decades, two major realisations emerged: 1. Binary interactions are highly common among massive stars  \citep{Sana2012, Kobulnicky2014}, and 2. Mass-loss rates at various stages \citep{Fullerton2006, Puls2006, Krticka2017, Bjorklund2020, Beasor2020} phases appear to be lower than previously thought. These two facts combined have led to a resurgence of the binary channel in forming WR stars. 

\section{The Wolf-Rayet binary fraction as a function of metallicity}

A fundamental parameter in this context is the WR binary fraction, which, for the Galaxy, is long-known to be $\approx 40\%$ \citep{Vanbeveren1998, vanderHucht2001}. However, this fraction is  not bias-corrected. A modern radial velocity survey for Galactic WR stars has been performed by \citet{Dsilva2020, Dsilva2022}, and in prep.  The observed binary fraction of WN stars remains comparable ($f_{\rm obs, WN} = 0.41\pm0.09$), and a bias correction of the WN population in the range $0 < \log P < 5\,$[d] implies that roughly half the Galactic WR stars are in binaries ($f_{\rm int, WN} = 0.52 \pm 0.13$).  However, for the putative decendents of WN stars -- WC stars -- an observed binary fraction of $f_{\rm obs, WC} = 0.58 \pm 0.14$ and an intrinsic fraction of $f_{\rm int, WC} > 0.74$ are derived. It is important to note that these fractions assume an underlying mass distribution for the companions: it is possible that all "apparently-single" WR stars have low mass companions ($M_2 \lesssim 5\,M_\odot$), if such companions are over-represented in nature. Moreover, the period distributions of the WN and WC populations are found to be distinct: Galactic WN binaries tend to exhibit periods of the order of days to weeks, while WC binaries tend to exhibits periods of months to many years (see Dsilva et al.\ in these proceedings). These striking differences  force us to reconsider the evolutionary connection between WN and WC stars, potentially implying that short-period WN binaries end up merging \citep{Dsilva2022}.

Binaries are thought to be especially critical at low metallicity ($Z$), where the winds are generally expected and observed to be even weaker. To investigate this, \citet{Bartzakos2001}, \citet{Foellmi2003SMC, Foellmi2003}, and \citet{Schnurr2008} established the binary fraction of WR stars in the SMC and LMC, which probe metallicity contents of $\approx 1/5\,Z_\odot$ and $\approx 1/2\,Z_\odot$, respectively. The results appeared rather surprising, since the the observed binary fraction were found to be relatively $Z$-independent ($\approx 30-40\%$). This is also observed in other galaxies such as M31 and M33 \citep{Neugent2014}.

\begin{figure}
\begin{center}
 \includegraphics[width=1\textwidth]{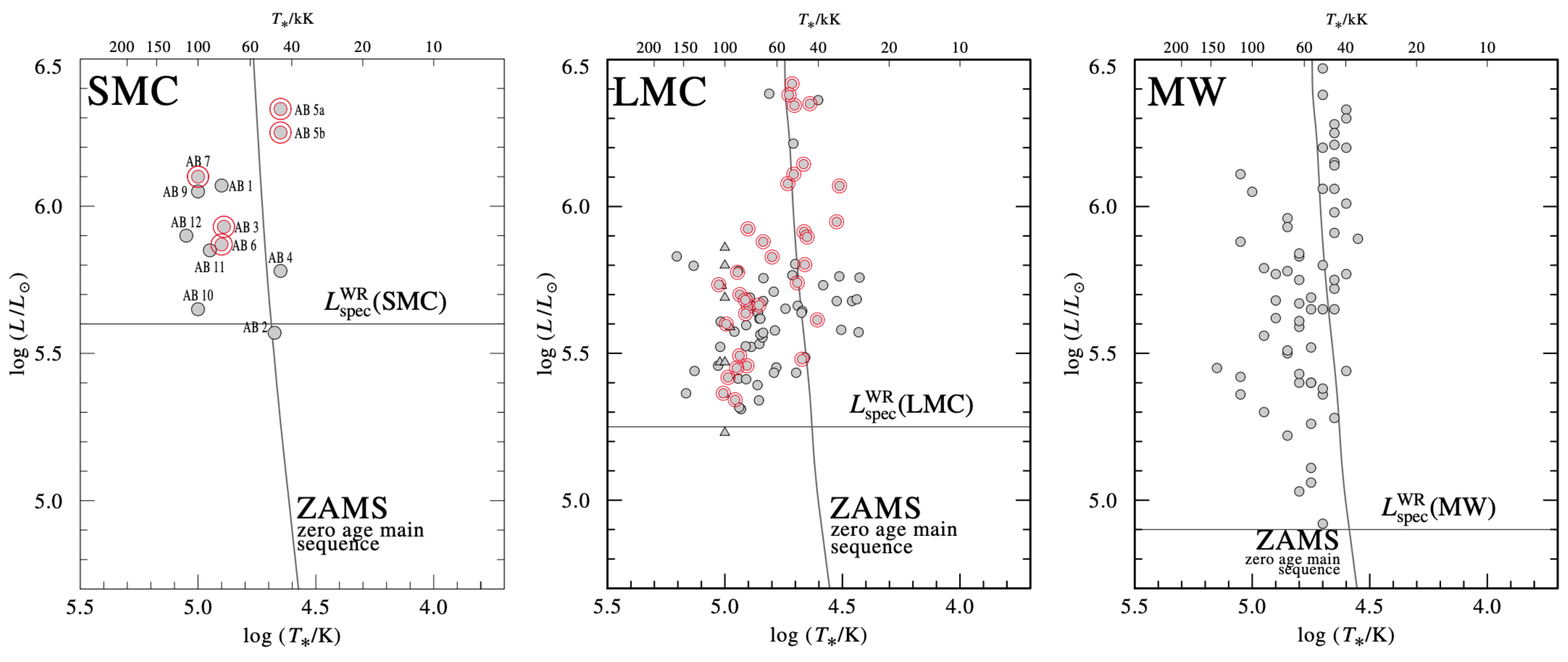} 
 \caption{The positions of the WN populations of the SMC, LMC, and Galaxy, adapted from \citet{Shenar_2020}. Red circled symbols are binaries. The Wolf-Rayet phenomenon stops below a $Z$-dependent luminosity threshold: stripped stars below this threshold would not exhibit the necessary winds to be classified as WR stars. The thresholds correspond to initial masses of $\approx 20\,M_\odot$, $\approx  25\,M_\odot$ and $\approx 40\,M_\odot$ for the Galaxy, LMC, and SMC, respectively. }
  \label{fig:WNHRD}
\end{center}
\end{figure}

A potential explanation for this result arises when one considers that fact that WR stars comprise a spectral class of stars with strong winds. But at low metallicity, not only is it harder to strip stars via winds, but it is also harder for the stripped stars to drive a wind. Hence, a rough threshold initial mass exists below which the stripped product would not appear as a WR star. Owing to comprehensive analyses of apparently-single and binary WR stars by \citet{Hamann2006, Hamann2019}, \citet{Hainich2014, Hainich2015}, and \citet{Shenar2016, Shenar2019} in the Milky Way, LMC, and SMC, these mass thresholds can be estimated empirically (Fig.\,\ref{fig:WNHRD}). Hence, \citet{Shenar2020WR} estimate that in the Milky Way, stars initially less massive than $\approx 20\,M_\odot$ would not appear as WR stars after stripping, while in the SMC, this limit grows almost to $\approx 40\,M_\odot$ \citep[see also][]{Aguilera-Dena2022}. This means that, at low $Z$, neither of the WR formation channel is effective, and it is not obvious that a metallicity trend should be present.

\section{"Single" WR stars: truly single, or binary-interaction products?}

In spite of the many advances, a major problem remains: many cWR stars do not have any identified companions. In fact, for the LMC population (Fig.\,\ref{fig:LMC_HRD}), this population of apparently-single WR stars comprises over 70\% of the entire cWR population. 

One possibility is that these stars are not truly single, but rather binary. The companions are  unlikely to be very massive or bright (give or take a few exceptions), because they would have been identified in previous surveys, unless they are so far to have anyhow avoided interaction with the WR component ($\log P \gtrsim 3.5\,$[d]). Rather, if they are there, they are likely low-mass and faint (or even completely dark: black holes). While stable mass transfer does not seem to be feasible in this case, stripping via common-envelope evolution might be. This was proposed by \citet{Schootemeijer2018}, who are now actively searching for hidden companions among the SMC WR population (PI: Schootemeijer, ESO program ID: 108.22M1). A different campaign aims to hunt for companions among the LMC WC population (PI: Shenar, ESO program ID: 105.207A). 

Another scenario related to binary interaction, which is however  harder to probe, is that the apparently-single WR stars were affected by binary interaction, but are now truly single. Mergers is one option, though why merging should lead to the formation of a WR star is not clear. Another option is that they are mass gainers in distrupted binaries (e.g., due to a supernova kick of the mass donor), and the resulting rapid rotation pushes them to evolve homogeneously \citep{Maeder1987}. Yet another option is that they were disrupted from the binary after stripping via third-body interactions \citep{Toonen2020}. These ideas need to be tested with modeling (e.g., population syntheses) and observations (e.g., proper motion and isolation of the targets) to investigate whether they can truly explain such a large part of the WR population.

However, it is also possible that single stars in the relevant mass regime ($M_{\rm ini} \gtrsim 25\,M_\odot$ in the LMC, \citealt{Shenar_2020}) do actually evolve into the WR phase naturally. One important mass-loss agent is represented by LBVs and their mass eruptions, which are still heavily debated \citep{Justham2014, Smith2015, Humphreys2016, Mahy2022}. It is also important to realise  the mass-loss of evolved stars with initial masses above $\approx 30\,M_\odot$ is not well constrained, since they generally do not become cool supergiants (the Humphreys-Davidson limit, \citealt{Humphreys1978}). Hence, red supergiant mass-loss has very little to do with WR stars, except perhaps for the few most luminuos red supergiants. In fact, some of the measured mass-loss rates derived for the most luminous red supergiants do reach the levels necessary for self-striping \citep{vanLoon2005, Beasor2022}, notably WOH~G64. 

Finally, increased mixing spurs bluewards evolution, whether induced by rotation \citep{Ekstroem2012} or processes like convection \citep{Higgins2019, Ramachandran2019}. Enhanced mixing was recently proposed by \citet{Gilkis2021} as a potential underlying agent for the Hupmhreys-Davidson limit, which appears to be $Z$-independent \citep{Davies2018, Sabhahit2021}. If true, this could also open a potential resolution for explaining the production of apparently-single cWR stars from a single-star prespective.


\bibliographystyle{apj}
\bibliography{./References}

\section{Q\&A}

\vspace{0.5cm}

\noindent
{\sc Dany Vanbeveren:} Can it be that we are still missing WR binaries in surveys? Are there any biases involved?

\vspace{0.5cm}

\noindent
{\sc Tomer Shenar:} Surveys of WR stars in the Magellanic Clouds are claimed to be virtually complete. However, I do not think that this accounts for WR stars with lines that are strongly diluted by a bright companion. It is hard to estimate the corresponding correction, since the frequency of such binaries depends on uncertainties in binary evolution. However, since WR stars per definition exhibit substantial emission in their spectra, I would be surprised if we are missing a large population of WR binaries.

\vspace{0.5cm}

\noindent
{\sc Sally Oey:} One clue to the binary nature of Wolf-Rayet stars is whether they are field stars. From what I recall from studies by Nathan Smith, Wolf-Rayet stars are not very "fieldy", unlike for example Oe/Be stars. This may suggest that single-star channels for some of the WR stars are relevant. 

\vspace{0.5cm}

\noindent
{\sc Raphael Hirschi:} I would also provide support the idea of underestimated mixing. Most of the studies adopt a constant mixing, but models show that mixing processes such as overshooting may be stronger for high-mass stars.

\vspace{0.5cm}

\noindent
{\sc Tomer Shenar:} I agree with both comments

\vspace{0.5cm}

\noindent
{\sc Koushik Sen:} What would be the mass ratio need to be for you to find the O-type companion in a WR binary?

\vspace{0.5cm}

\noindent
{\sc Tomer Shenar:} This depends on the method. If you are relying on the radial velocity measurements of the Wolf-Rayet component, then in principle the mass ratio does not matter, just the amplitude of the Wolf-Rayet component. But if you want to identify the O-type star in the spectrum, then it would depend on a lot of parameters, and naturally the data quality. However, from experience, I would argue that it would generally be difficult to hide an O-type star in the spectrum.

\end{document}